\documentclass[epj]{svjour}

\usepackage{latexsym} 
\begin{document} 
\title{Symmetry in the equations of nonlinear dynamics of semiconductor laser subject to delayed optical feedback}
\titlerunning{Symmetry in the equations of nonlinear dynamics\ldots}
\author{A.\,P.\,Napartovich \and 
        A.\,G.\,Sukharev \/\thanks{e-mail: sure@triniti.ru}}
\institute{State research center Troitsk Institute for Innovation and Fusion Research, 142190 Troitsk, Moscow reg., Russia}
\date{Received: 8 June 2011 / Revised version: *} 
\abstract{ 
Delayed feedback laser dynamics is described by means of Lang-Kobayashi equation model. Since a lot of initial states asymptotically approach to periodic attractor in the phase space, only periodic steady-state regimes have been studied here. Lyapunov transformation allows us to reduce problem to the differential equation of the first order whereas the spectrum of laser oscillation is governed by the appropriate eigen value problem. Using the symmetry proves that there is the dual laser system with anticipated feedback which has the same dynamic characteristics as laser with delayed feedback.
}  %end of abstract 
\maketitle
\section{Introduction}
\label{intro} 
A delayed optical feedback (DOFB) semiconductor laser is characterized by a variety of dynamical regimes~\cite{bib:KL}, some of them being used in optical communication systems. The complexity of these regimes is caused by the presence of two coupled resonators, one of which is formed by the reflecting end facets of a semiconductor crystal, and the other - by a combination of a highly reflecting end facet of the crystal and an external mirror. Lang and Kobayashi~\cite{bib:LK} derived dynamic equations for a laser diode with an additional external mirror by neglecting the reflection of external radiation from the diode facet. The Lang - Kobayashi (LK) equations form the basis of modern theories describing the dynamics of DOFB laser diodes. The system of LK equations contains six key parameters, changing of which results in origin of a variety of dynamical regimes up to chaos. There are two alternative methods to produce a variety of dynamical regimes in the semiconductor laser system without external mirror. In the first of them an optically injected diode laser~\cite{bib:WKL,bib:KGG} is used. In the second one a system of the mutual optically coupled diode lasers~\cite{bib:RW,bib:KPG,bib:PLGK} is applied. There is a common element which has all of these lasers. The part of the radiation is injected into laser resonator from the outside. It may be harmonic radiation as in~\cite{bib:WKL,bib:KGG}, or undefined field in the general case. 

From mentioned point of view a delayed feedback laser and a system of the mutual optically coupled lasers are almost equivalent. The reflection of the field from the mirror can be interpreted as an existence of the some radiating element behind the mirror. So some virtual laser placed in image space exchanges radiation with the original laser. The characteristics of virtual laser and mutual coupling constants are unknown and must be defined with the help of the theory.  This problem is relatively simple if the laser regime is stationary. Yet it is well known that between this state and chaos, the delayed feedback laser demonstrates a lot of dynamical regimes. All of them belong to either transient states or to steady-state periodic~\cite{bib:NS} (conditionally-periodic~\cite{bib:SN}) oscillations. The existence of nontrivial dynamical properties of the nonlinear system is caused by delayed feedback. The new dynamic state of the system is appeared at a bifurcation points, sequence of which is developed with feedback variation. Asymptotic behavior of the dynamic state of the laser system is one of the steady-state regular oscillations, if the regime is not chaotic. The steady-state oscillations are reached in the inversion relaxation time. That is why except the short transient states the solving of the nonlinear system can be studied as a series expansion of the periodic function. It is natural that a numerical integration of LK equations is simple enough in comparison with a solving of this expansion. But when DOFB laser can have several solutions, an infinite set of initial conditions must be verified to find some of them. Conversely, the our method can be represented as a problem of eigen values, each being arranged by  the energy stored in oscillations. To solve the nonlinear LK equations along periodic path it is used Lyapunov's transform~\cite{bib:G}. For that it is necessary to linearize system of equations with the help of small variations procedure along the closed path. Separating the tangent variations we find a new nonlinear equation with the same temporal evolution as initial problem. By applying mirror symmetry a theory permits to find parameters of virtual laser located in the image space. The dual laser in image space has anticipated dynamic with respect to original. So used concept couples a dynamics of DOFB laser with proper dynamics of the mutual optically coupled lasers.   

\section{Model of nonlinear laser dynamcis}
\label{sec:1} 

The nonlinear dynamics of the semiconductor laser with delayed optical feedback (DOFB) arising from the reflection of the part of the radiation from external mirror is described by Lang-Kobayashi (LK) equations~\cite{bib:LK}: 
\begin{eqnarray}
\frac{{\partial E}}{{\partial t}} & = & (1 - iR)NE + iM{e^{i\kappa }}E(t - {\tau _D}), \nonumber \\
\label{eq:1} \\ 
T\frac{{\partial N}}{{\partial t}} & = & P - N - (1 + 2N){\left| E \right|^2}.
\nonumber 
\end{eqnarray}

The first equation is used for description of the slowly varying envelope of the field amplitude $E(t)$, the second one describes the inversion dynamics $N(t)$ in relation to dimensionless time. $R$  is the linewidth enhancement factor being responsible for reflective index changes with increasing gain. The module and phase of the feedback is denoted as $M$ and $\kappa$ accordingly. The time lag of the argument in the delayed term of the equation is defined by the transit time over external resonator ${\tau _D}$. The field gain is proportional to population inversion, the dynamics of which is described by the second LK equation~\cite{bib:RW,bib:AD}. Dimensionless value $N$ is expressed by means of carrier concentration $N_c$ over threshold ${N_{th}}$: $N = {1/2}g{\tau _{ph}} ({N_c} - {N_{th}})$, where $g=(c/n)\partial G/\partial N_c$ is a differential gain of the active medium. Here $c$ and $n$ are the speed of light and group refraction index. The photon lifetime is $\tau _{ph}^{ - 1} = {{(c} / n})({L^{ - 1}}\ln {r^{ - 1}} + {\alpha _c})$, while ${\alpha _c}$ is distributed losses in passive structures, diode cavity length equals $L$ and amplitude reflectivity of the end facets corresponds to $r$. The threshold carrier density ${N_{th}} = {N_{tr}} + {(g{\tau _{ph}})^{-  1}}$ takes into account active losses and passive losses of the radiation inside volume and passive losses on the chip facets. The carrier density for transparency $N_{tr}$ is used for denotation of the active losses. Here $P = {1/2}g{\tau _{ph}}(j{\tau _s} - {N_{th}})$   is normalized pump intensity, and ${\tau _s}$  is a carrier lifetime without induced transitions, $j$  is a constant rate at which carriers are injected into active region. Dimensionless time is implemented with the help of the normalization to value photon lifetime ${\tau _{ph}}$,  which accounts for both passive in-cavity losses and losses on diode facets. In particular, the dimensionless inversion relaxation time is defined as: $T = {{\tau _s} / {\tau _{ph}}}$. The typical value for this parameter is $T{\approx} 1000$. The intensity of the field can be evaluated from the dimensionless $E$ as a photon flux per unit area $2cE^2/ng\tau_s$  with energy $\hbar\omega$ per photon.

The LK equations system is provided with memory about previous states. As a result à wide variety of different states~\cite{bib:KL} can be found as a function of initial distribution of the field and population inversion over time interval equaling time delay. Nevertheless, when pump level is small enough, randomly chosen set of initial conditions lead to regular dynamics. Based on  theoretical grounds of the Kolmogorov-Arnold-Moser theory~\cite{bib:K} if the perturbation formed  by delayed feedback is small enough, most of the phase trajectories of the system belong to the periodic invariant torus. Therefore it is very important to investigate the steady-state periodic oscillations for LK equations. The mathematical aspects of the theory of equations with delayed feedback are under construction~\cite{bib:VL}.  

In this work the regular periodic steady-state solutions of the LK equations are studied with the help of some methods of the linear theory of perturbations. When periodic oscillations are steady-state, the perturbations are described by linear differential equations with periodic coefficient. The system is complicated, but the certain equivalent system ordinary equations can be found from the fact of solution periodicity. The reducibility transformation was proved by Lyapunov by using the property of the solution periodicity only~\cite{bib:G}. This method helps us to consider the delayed feedback for periodic steady-state modes as some set of the effective DOFB constants. The combination of linear problem for perturbations and equations for effective coefficients of DOFB permits us to solve the inverse problem of finding nonlinear oscillations for LK equations.

After substitution $E = {E_s}\exp (i\beta t + \psi (t))$ the dynamic of the field can be studied with the help of complex phase function $\psi$.  Unknown parameters were eliminated in accordance with stationary state: $\frac{{\partial N}}{{\partial t}}=0,$ $\frac{{\partial E}}{{\partial t}} = i\beta $, where  $\beta  = M(\cos \chi  + R\sin \chi )$ and $\chi  \equiv \beta {\tau _D} - \kappa $ which is a new representation of the feedback phase. The stationary value of the population inversion  ${N_0} =  - M\sin \chi $ defines the filed intensity $E_s^2 = (P - {N_0})/(1 + 2{N_0})$ in the stationary regime,  and $\beta $ is a frequency detuning from the center of the reference wave. If the feedback strength is small enough (and hence ${N_0} \ll 1$), the equations for variables $\psi $ and $n=N-{N_0}$  are:

\begin{equation} 
\frac{\partial }{{\partial t}}\psi  = (1 - iR)n + iM{e^{ - i\chi }}\left[ {\exp (\psi (t - {\tau _D}) - \psi (t)) - 1} \right]  \label{eq:2}
\end{equation}
\begin{eqnarray}
\frac{{\partial n}}{{\partial t}} =  - \frac{n}{{{T_1}}} - \frac{{\omega _r^2}}{2}(\exp (2{\mathop{\rm Re}\nolimits} \psi ) - 1)\nonumber\end{eqnarray}
Here we have used the following definitions: $\omega _r^2 = 2(P - {N_0})/T \approx 2P/T$ and ${T_1}^{ - 1} = {T^{ - 1}}(1 + 2P)/(1 + 2{N_0}) \approx {T^{ - 1}}$. 
To study the behavior of the perturbations to the solution of the system~(\ref{eq:2}) it is important to take into consideration that usually ${\omega _r} \gg T^{ - 1}$, that is, the characteristic frequency of oscillations is much higher than inverse time of the inversion relaxation. So our description is bounded only to those frequencies, which obey this condition. 

\section{Perturbation analyze of LK equations}
\label{sec:2} 

To analyze the features of the equations ~(\ref{eq:2}) one can perform a transfer to linear equations for small perturbations. With the help of substitution $\psi  \to \psi  + \delta \psi ,\quad n \to n + \delta n$ from initial nonlinear equations it is easy to produce linear system of equations for small variances from exact solution. To study system of equation for variances from periodic solution, we can consider all coefficients of such system as being formed from periodic functions. With the help of Lypunov transformation $X=L(t)Y$ this system $\dot X=P(t) X$ can be reduced to the system of the equation with constant coefficients. In the one-dimensional case the reduced system is~\cite{bib:G}:
$$\dot Y = i\omega Y.$$
This equation corresponds 
$$\dot\Phi =  \dot X/iX = \omega + \dot L/iL,$$
for original variable $X$. Because of the Lyapunov transformation $L(t)$ is periodic function by definition, the deformed limit cycle in this space is governed by the nonlinear equation:
$$\dot\Phi =  \omega + \sum s_k \exp{(ik\Phi)},$$  
where $s_k$ is some constant, obtained as result of series expansion procedure for the periodic function $\dot L/L$. So the Lyapunov transformation defines new nonlinear equation, which has the same temporal evolution as the original system. The next calculations are directed to determine the specific form of the expansion constants.

Without DOFB the system of equations~(\ref{eq:2}) has stable stationary state $\psi  = 0$, so any initial perturbation decreases with time.  DOFB changes the behavior of the system and this causes appreciable structural reconfiguration of the phase-plane portrait. The different types of the solutions are divided among them by the bifurcation points.

To solve the linear system equations for small perturbations produced from~(\ref{eq:2}) we divide the system into two different parts: the homogeneous part $\bf A$ and inhomogeneous part $\bf B$ with delayed feedback. The first one doesn't have the feedback term and can be easily solved, the second part is governed only by DOFB. The strong reconfiguration of the phase-plane portrait shows us that any attempt to find the solution as a finite expansion in series of powers of feedback coefficient $M$ is incorrect. Furthermore the homogeneous part of the solution is vanishing value as opposed to a specific solution. The specific solution of this system can be produced from the homogeneous part of the solution with the help of constant variation method~\cite{bib:VLK}. Suppose that $\delta \psi  = \delta x + i\;\delta y,\;\delta {\psi ^ * } = \delta x - i\;\delta y$ and $\delta \xi=\delta \psi (t  - {\tau _D}) - \delta \psi (t),\; \delta {\xi ^ *}=\delta {\psi ^ *}(t  - {\tau _D}) - \delta {\psi ^ *}(t)$, $\Lambda(t) = M\exp ( - i\chi  + \psi (t  - {\tau _D}) - \psi (t ))$, then: 
\begin{equation} 
\left( {\begin{array}{*{20}{c}}
   {\delta \psi (\tau )}  \\
   {\delta {\psi ^ * }(\tau )}  \\
   {\delta n(\tau )}  \\
\end{array}} \right) = \int\limits_{}^\tau  {dt} \; 
{\mathord{\buildrel{\lower3pt\hbox{$\scriptscriptstyle\frown$}}
 \over {\mathit T}}}
(\,\exp\, (\int\limits_t^\tau  {dt'} {\bf A}))\,
\left( {\begin{array}{*{20}{c}}
   {i\Lambda(t)}\,{\delta \xi}  \\
   { - i{\Lambda^ * }(t)}\,{\delta {\xi ^ *}}  \\
   0  \\
\end{array}} 
\right)\label{eq:3}  
\end{equation}
here ${\mathord{\buildrel{\lower3pt\hbox{$\scriptscriptstyle\frown$}}
 \over {\mathit T}}} (\exp )$ is a chronological exponent~\cite{bib:BSH}, an external integral in formula~(\ref{eq:3}) is an indefinite one.The matrix ${\bf A}$ can be written as:
 $${\bf A} = \left( {\begin{array}{*{20}{c}}
   0 & 0 & {1 - iR}  \\
   0 & 0 & {1 + iR}  \\
   { - (\omega _r^2/2){{\mathop{\rm e}\nolimits} ^{2u}}} & 
   { - (\omega _r^2/2){{\mathop{\rm e}\nolimits} ^{2u}}} & 
   { - 1/T}  \\
\end{array}} \right),$$
where the new symbolic designation have been introduced:
$$2u = (\psi  + {\psi ^ * }). \;$$

As the diagonal term in the matrix ${\bf A}$  is proportional to $ - 1/T$, the homogeneous part of the solution relaxes to zero with time. The only specific solution of the inhomogeneous system defines the dynamics at relatively long time intervals. That is why the behavior close to periodic attractor is defined by the integration interval in the vicinity of upper limit, when $t \to \tau $. 

The features of the solution are governed by the spectrum of the matrix ${\bf A}$. It should be taken into account that only two eigen values have time dependence: ${\lambda _0} = 0,\; {\lambda _{1,2}} =  - 1/2T \pm i{\omega _r}\exp (u)$. So three eigen vectors of the matrix ${\bf A}$ can be expressed:

$\begin{array}{ll}
{v_0} = \left( {\begin{array}{{l}}
   i  \\
   { - i}  \\
   0  \\
\end{array}} \right),\quad {v_{1,2}} = \left( {\begin{array}{{l}}
   {(1 - iR)}  \\
   {(1 + iR)}  \\
   {{\lambda _{1,2}}}  \\
\end{array}} \right)
\end{array}$\linebreak

In three dimensional complex eigen vector space of the matrix ${\bf A}$ it is possible to find two dimensional real subspace ${v_0},\;{v_1} + {v_2}$ in which the third line ($\delta n$ component) is zero if oscillation frequency is greater than relaxation rate ${\omega _r} \gg  T^{-1} $. The choice like this results from the absence of feedback via population inversion, so the third line of the perturbation vector is filled with zeros. With the help of two-dimensional subspace one can exclude dependence on variable $\delta n$ and decrease the dimension of the equation system. The exterior integral in matrix equation~(\ref{eq:3}) implies integration in small vicinity of the variable $\tau$.  Therefore the inner integral located under the exponent is taken on infinitely small interval $\left(t,\;\tau \right)$. Every integration step $h$ on this interval doesn't change vector ${v_0}$, but changes vectors ${v_1}, {v_2}$. Nevertheless their sum ${v_1} + {v_2}$ is again the sum of eigen vectors  for the next step of integration with an accuracy $o(h)$. This specific property corresponds to adiabatic approximation. If we define the primitive function $\Phi $ for $\exp(u)$ function as:
\begin{equation}
\int\limits_t^\tau  {dt'} \exp (u) = \\ \int\limits_t^\tau  {dt'} \;\exp ((\psi  + {\psi ^ * })/2) = \Phi (\tau ) - \Phi (t) \label{eq:4}
\end{equation}
and retain the same order of accuracy to chronological operator in  equation~(\ref{eq:3}), then:

$\begin{array}{l}
{\mathord{\buildrel{\lower3pt\hbox{$\scriptscriptstyle\frown$}} 
\over {\mathit T}}} 
(\,\exp\, (\int\limits_t^\tau {dt'} A\,))({v_1} + {v_2}) = \\ {{\mathop{\rm e}\nolimits} ^{ + i{\omega _r}(\Phi (\tau ) - \Phi (t))}}{v_1}(\tau ) + {{\mathop{\rm e}\nolimits} ^{ - i{\omega _r}(\Phi (\tau ) - \Phi (t))}}{v_2}(\tau )
\end{array}$

If periodic oscillations are steady-state, the system of linear differential equations for small perturbations will be analogous to differential equations with periodic coefficients. In this case the system can be reducible from the viewpoint of Lyapunov ~\cite{bib:G}. In order to implement Lyapunov's transformation to the system of ordinary equations, the expression  $M\exp ( - i\chi  + \psi (\tau  - {\tau _D}) - \psi (\tau ))$ can be expanded in the $\exp (ik\omega \Phi )$  function series, where numbers $k$ are whole numbers. This system of the functions is complete system, since $\exp (ik\omega \Phi )= \left[ \exp (i\omega \Phi )\right]^{k}$ forms power series, but it is not orthogonal. 
\begin{equation}
\begin{array}{c}
\Lambda(t) = M\exp ( - i\chi  + \psi (t  - {\tau _D}) - \psi (t)) 
=\\ =\sum\limits_{k =  - \infty }^\infty  {{h_k}} \exp (ik\omega \Phi ),
\end{array} \label{eq:5}
\end{equation}
${h_k}$ - are the effective feedback coefficients for harmonics with the number   $k$. So the two equations system can be developed from the matrix equation~(\ref{eq:3}) (it is clearly that $ \delta\Phi = \delta\Phi({\mathop{\rm Re}\nolimits} \psi)$ since $\ln (\dot \Phi ) = {\mathop{\rm Re}\nolimits} \psi $, $\delta \Lambda=\Lambda(t)\delta\xi\propto \delta\Phi$):
\begin{equation}
\begin{array}{c}
\left( {\begin{array}{{l}}
   {\delta \psi (\tau )}  \\
   {\delta {\psi ^ * }(\tau )}  \\
\end{array}} \right) = \int\limits_{}^\tau  {dt} \; \delta \Phi\; \times\\
\left[ {{q_ + }\left( {\begin{array}{{l}}
   {1 - iR}  \\
   {1 + iR}  \\
\end{array}} \right)\cos ({\omega _r}\Phi (\tau ) - {\omega _r}\Phi (t)) + 
{q_ - }\left( {\begin{array}{{l}}
   { + i}  \\
   { - i}  \\
\end{array}} \right)} \right],\\
{q_ + } =  - \omega\; {\sum {{[k(h_k+h^*_{-k})/2]}\exp (ik\omega \Phi(t) )}} ,\\
{q_ - } = R{q_ + } - \omega\; {\sum {{[k(h_k-h^*_{-k})/2i]}\exp (ik\omega \Phi(t)  )} } .
\end{array} \label{eq:6}
\end{equation}

\section{Nonlinear solution and mirror symmetry}
\label{sec:3} 

The system of equations~(\ref{eq:6}) is correct for any small variation towards nonlinear solution. If we take into consideration the variations $\delta \psi $ which are tangent to the nonlinear solution, $\delta \psi \propto \dot \psi $, then we can convert the equations~(\ref{eq:6}) to obtain an assumed solution of  nonlinear equations~(\ref{eq:2}). 
The motion along actual path produces the canonical transformations preserving the invariants. According to Noether's theorem~\cite{bib:K} a invariance relative to the translation along vector tangent to the cyclic phase path leads to the system has the first integral. Consequently perturbation $\delta \psi$ for established states should be searched proportional to $\dot \psi$.
The equations~(\ref{eq:6}) formally have the form $\dot{\delta \psi}  = f(\Phi )\delta \Phi ( {\mathop{\rm Re}\nolimits} \psi)$. Hence there is some nonlinear master equation $\dot \psi  = F(\Phi )$, from which it follows given equation for small variation while $\delta F/\delta \Phi = f(\Phi )$. Furthermore the differentiation of this nonlinear equation produces also the new expression $\ddot \psi  = f(\Phi )\dot \Phi  = f(\Phi )\dot\Phi({\mathop{\rm Re}\nolimits} \psi)$.        
The real part of the tangent variations ${\mathop{\rm Re}\nolimits} \delta \psi$ can be expressed in terms of $\Phi $ function which was introduced in~(\ref{eq:4}). Value of the $\Phi $ defines the phase of the nonlinear oscillations and its time dependence obeys the differential equation of second order: 
\begin{equation}
{\mathop{\rm Re}\nolimits} \dot \psi = 
\ddot \Phi /\dot \Phi  = - \sum\limits_k^{} {\frac{{{{(k\omega )}^2}}/2i}{{{{(k\omega )}^2} - \omega _r^2}}} ({h_k}-{h^*_{-k}})\exp (ik\omega \Phi) 
\label{eq:7}
\end{equation}

Here the $k$ numerates motion on different average frequencies ${k\omega \overline {\dot \Phi }}$. The number of elements under a summation sign specifies actual dimension of the motion equations system. The more the number of components is used, the more detailed nonlinear dynamics is described. The order of the equation~(\ref{eq:7}) can be decreased in the following way: by multiplying into the function  $\dot \Phi $ and performing the integration. As a result the  equation of the first order was derived (${s_{k}^{}}\equiv s_k({h_k},{h_{-k}^{*}})={s_{-k}^{*}}$):
\begin{equation}
\dot \Phi  - \sum\limits_{k }^{}  {s_k} \exp (ik\omega \Phi ) = E,
\quad {s_k} = \frac{1}{2} \frac{{(k\omega )({h_k} - h_{ - k}^*)}}
{{{{(k\omega )}^2} - \omega _r^2}} \label{eq:8}
\end{equation}
In this equation the integration constant $E$ appeared. The new condition to exclude this constant can be added. From the system of equations~(\ref{eq:2})  it follows that $\overline {\exp (2{\mathop{\rm Re}\nolimits} \psi)} - 1 = 0$ with accuracy of the small parameter $M/P$. The constant $E$ fulfils the role of the energy level, and its meaning can be found from the expression~(\ref{eq:8}) with the help of condition $\overline {{{\dot \Phi }^2}}  = 1$. The converted condition is $E\overline {\dot \Phi }  = 1$. The period of the oscillation repetition can be determined via the method of the separation of variables: 
\begin{equation}
\frac{1}{2\pi} \int_0^{2\pi } {d\Theta } {\left( {E + \sum\limits_{k }^{} {s_k} \exp (ik\Theta)} \right)^{ - 1}} =  \omega /\Omega  = E
\label{eq:9}
\end{equation}

So, Lyapunov's transformation is fulfilled with the help of the $\dot L/L$ function expansion. In the limit of $1/T\to 0$, this transformation results in integrability of the system. The constant $\omega $ is an eigen number, and pulse repetition frequency equals $\Omega  =\overline{\dot{\Theta}} = \omega \overline {\dot \Phi }$.

Thus,  the complex amplitude of oscillation $\psi $ is evaluated in terms of the integral:
\begin{equation}
 \psi (\tau ) = i \int {dt} \quad  \sum {d_k} \exp (ik\omega \Phi (t) ),
\label{eq:10}\end{equation}
\begin{equation}
{d_k}=\frac{1}{2}\left\{ {2{h_k} + ({h_k}-{h_{ - k}^*}) \frac{{{\omega _r^2}
 (1 - iR)}}{{{{(k\omega )}^2} - \omega _r^2}} } \right\}
\label{eq:11}\end{equation}

To define sequence of the constants ${h_k}$  in expansion of the $\Lambda(t)$ function we introduce a basis function set ${\bf P_{t,k}} = \exp (ik\omega \Phi (t))$. The conclusion is used here that all of them have the basic period of oscillation $2\pi /\Omega$, while $k$ is the harmonic number. To build the search algorithm of the constants ${h_k} $  we apply the method of Moore-Penrose pseudo inverse matrix~\cite{bib:G}. Then a vector with components ${h_k}$ equals to:  
\begin{equation}
h = {({\bf P^ + }\bf P)^{ - 1}}({\bf P^ + }\Lambda(t)),\quad {\bf P^ + } = {({\bf P^*})^T}\label{eq:12}
\end{equation}

Vector of the coefficients $h$ represented by the formula~(\ref{eq:12}) is the best approximation in terms of least-squares if a finite number of basis functions are used.  The matrix ${({\bf P^ +}{\bf P})_{mk}}$ is the nonsingular Hermitian matrix and its elements $(2\pi /\Omega ){f_{k - m}}$ are specified by  Fourier integral over period:

${f_{k - m}} = (\Omega /2\pi )\int\limits_{ - \pi /\Omega }^{\pi /\Omega } {dt} \exp (i\omega (k - m)\Phi (t)) = {f_{m - k}^*}$.
 
The value of the matrix element is defined by difference kernel, so any diagonal contains single element in all places. It was found that their meaning can be calculated directly from equation~(\ref{eq:8}). If all values of the  $({h_k} -h_{-k}^*)$ are real numbers, the phase function has the reflection symmetry: $\Phi (t) =  - \Phi (-t)$. In general case it is possible to construct the correspondence  $\Phi '(t) = - \Phi (-t)$. This dual function is again the solution of the equation~(\ref{eq:8}), and yet the numbers ${s'_k} \equiv s_k(-h_{-k},-h_k^*) = s_k^*$ have been modified. The dual system solution is spectral-symmetric to original one. In accordance with (\ref{eq:5},\ref{eq:12}) effective feedback coefficients ${h_n}$  satisfy the following equation:
\begin{eqnarray}
{h_n} = M{e^{ - i\chi }} \frac{\omega }{{2\pi }}  
\int\limits_{ - \pi /\Omega }^{\pi /\Omega } {d\tau } \; \dot\Phi(\tau)
\exp \left( { - in\omega \Phi(\tau) }\right) \times \nonumber\\ 
\exp \left( {  + i \int\limits_\tau ^{\tau-\, {\tau_D}} {dt\;
\sum {{d_k}}\exp (ik\omega \Phi(t) )} } \right),\label{eq:13}
\end{eqnarray}
here the  inverse matrix factor ${({\bf P^+}{\bf P})^{ - 1}_{nk}} = ({\omega }/{{2\pi })}{s_{n - k}}$ was properly taken into account via $\dot\Phi$ and with the help of the equation~(\ref{eq:8}). Leading diagonal of this matrix is filled with numbers ${s_0} = E$, in the other cases numbers ${s_k}$ which were defined earlier in formula~(\ref{eq:8}) under $k \ne 0$ are used. It is obvious that elements ${(2\pi /\Omega)f_{k - n}}$ of the matrix ${{({\bf P^+}{\bf P})}_{nk}}$ can be produced by solving linear system: ${s_{k }}{f_{ k - n}} =  ({\Omega}/{\omega}){\delta _{0n}}$, here ${\delta _{mn}}$ is the unit matrix and ${E^{ - 1}} = \Omega /\omega $. And finally coefficients  ${d_k} \equiv d_k({h_k},{h_{-k}^{*}})$ used in last equation~(\ref{eq:13}) are introduced in~(\ref{eq:11}). 

To transform into dual system we symmetric reflect the index sign in~(\ref{eq:13})  and the direction of the time axis should be changed too.  As result the sign of the time delay ${\tau _D}$ is changed:  
\begin{eqnarray}
{h_{-n}} =   M{e^{ - i\chi }} \frac{\omega }{{2\pi }}
\int\limits_{ - \pi /\Omega }^{\pi /\Omega } {d\tau } \; \dot\Phi '(\tau) 
\exp \left( { - in\omega \Phi '(\tau) }\right) \times \nonumber\\ 
\exp \left( { - i \int\limits_\tau ^{\tau  + {\tau _D}} {dt\;
\sum {{d_{-k}}}\exp (ik\omega \Phi '(t) )} } \right) \label{eq:14}
\end{eqnarray}
So the transformation $ - {h_{ - n}} \to {{h'}_n}$ and $ \chi \to \chi '$ (together with ${-d_{ - k}} \to {{d\,'}_k} \equiv d_k({{h'}_k},{{h'}_{-k}^{*}})$, ${{s'}_k} \equiv s_k({h'}_{k},{h'}_{-k}^*)$)  sets up a correspondence with the vector of the dual system.  Then dual system is governed by the same equation~(\ref{eq:13}) only with opposite sign of the delay time  ${-\tau _D}$. Having compared the expressions~(\ref{eq:13}, \ref{eq:14}), the  necessary condition for the feedback coefficient phase  $\chi ' - \chi  = \pi $ or  $\kappa+\kappa' = 2 \,\arctan(1/R)$ can be found. If this condition is realized, the dynamics of the both systems have coincident period of oscillation $ 2\pi/\Omega $. It follows from the mirror symmetry which is defined here as the time axis reflection and mirror inversion of the spectral components $ - {h_{ - n}} \to {{h'}_n}$. So both systems have the same eigen number $\omega $, but feedback delay time transforms to feedback anticipate time. Therefore the effect of the delayed feedback can be interpreted as the mutual optical coupling between the laser and its virtual image. However coefficients of the mutual optical coupling have difference in phase. So, the dynamics of the laser with delayed optical feedback and response of proper pair ($\psi$,$\psi'$) of the optical coupled lasers are equivalent, while the regimes of the periodic oscillations are used. The time lag, equaled to time delay for DOFB laser, retains between coupled lasers with dual dynamics.
\begin{eqnarray} 
\frac{\partial }{{\partial t}}\psi  = (1 - iR)n + iM{e^{ - i\chi }}\left[ {\exp (\psi' - \psi) - 1} \right]  \nonumber \\
\label{eq:15} \\
\frac{\partial }{{\partial t}}\psi'  = (1 - iR)n' - iM{e^{ - i\chi }}\left[ {\exp (\psi - \psi') - 1} \right]  \nonumber
\end{eqnarray}

\section{Conclusions}
\label{sec:4} 

For LK equations we have developed new spectral method of the solving of the nonlinear dynamic states. Without direct integration of original LK equations it is possible to analyze the periodic steady-state modes of diode laser with delayed feedback. It was found that the periodic attractor evolution is described with the first order differential equation. This description was introduced with the help of Lyapunov's theoretical concepts about reducibility of the linear differential equations with periodic coefficients to ordinary one. The constants in this equation are defined via spectrum decomposition of the effective delayed feedback. Calculation of these coefficients is a nonlinear algebraic problem. The algebraic equations are developed via the Fourier transform with an anharmonic oscillation phase, which obeys the above-mentioned differential equation of the first order. Finally it was shown that for any DOFB laser the dual laser system with anticipated feedback can be found with the help of symmetry principle. The optically coupled laser and its dual image produce the same dynamics as a single DOFB laser.

\end{document}